# Evolution of Electronic Structure in Atomically Thin Sheets of WS$_2$ and WSe$_2$

*Weijie Zhao[a,c,#], Zohreh Ghorannevis[a,c,#], Leiqiang Chu[a,c], Minglin Toh[d], Christian Kloc[d], Ping-Heng Tan[e], Goki Eda[a,b,c,*]*

[a] Department of Physics, National University of Singapore, 2 Science Drive 3, Singapore 117542

[b] Department of Chemistry, National University of Singapore, 3 Science Drive 3, Singapore 117543

[c] Graphene Research Centre, National University of Singapore, 6 Science Drive 2, Singapore 117546

[d] School of Materials Science and Engineering, Nanyang Technological University, N4.1 Nanyang Avenue, Singapore 639798

[e] State Key Laboratory of Superlattices and Microstructures, Institute of Semiconductors, Chinese Academy of Sciences, Beijing 100083, China

[*] E-mail: g.eda@nus.edu.sg

[#] These authors contributed equally to this work.

**Abstract**

Geometrical confinement effect in exfoliated sheets of layered materials leads to significant evolution of energy dispersion with decreasing layer thickness. Molybdenum disulphide (MoS$_2$) was recently found to exhibit indirect to direct gap transition when the thickness is reduced to a single monolayer. This leads to remarkable enhancement in the photoluminescence efficiency, which opens up new opportunities for the optoelectronic applications of the material. Here we report differential reflectance and photoluminescence (PL) spectra of mono- to few-layer WS$_2$ and WSe$_2$ that indicate that the band structure of these materials undergoes similar indirect to direct transition when thinned to a single monolayer. Strong enhancement in PL quantum yield is observed for monoayer WS$_2$ and WSe$_2$ due to exciton recombination at the direct band edge. In contrast to natural MoS$_2$ crystals extensively used in recent studies, few-layer WS$_2$ and WSe$_2$ show comparatively strong indirect gap emission along with distinct direct gap hot electron emission, suggesting high quality of synthetic crystals prepared by chemical vapor transport method. Fine absorption and emission features and their thickness dependence suggest strong effect of Se p-orbitals on the d electron band structure as well as interlayer coupling in WSe$_2$.

Atomically thin sheets of $MoS_2$ consisting of a single to a few monolayers were recently found to exhibit a set of unusual properties associated with its two dimensional (2D) structure and high crystal quality, triggering significant research interest.[1-18] Similar to graphite, $MoS_2$ crystallizes in a van der Waals layered structure where each layer consists of a slab of S-Mo-S sandwich. Individual monolayers can be mechanically peeled off from a bulk crystal and isolated as graphene-like 2D crystal.[19] While $MoS_2$ is traditionally known as solid-state lubricant and catalyst, its monolayer crystal hold promise in novel electronic and optoelectronic applications as well as provide access to fundamental physical phenomena.

Electronically, a bulk $MoS_2$ is an indirect gap semiconductor with the bandgap in the near-infrared frequency range (~ 1.2 eV).[20] In contrast, monolayer $MoS_2$ is a direct gap semiconductor with the band gap in the visible frequency range (~ 1.9 eV).[4-5] Confinement of carriers in the out-of-plane direction induces gradual increase in the band gap with decreasing thickness.[21] The indirect to direct crossover occurs at the monolayer limit resulting in strong contrast in photoluminescence (PL) efficiency between single and multilayer sheets.[4-5] Remarkable electrical properties of mono- and few-layer $MoS_2$ are increasingly revealed in its field effect transistor characteristics such as exceptionally large on/off ratio, small subthreshold swing, ultra-low off-state energy dissipation, high field effect mobility, and large current carrying capacity.[1-3] Explicit inversion symmetry breaking in monolayers allows valley selective pumping of charges using circularly polarized light.[6-8] High electronic quality of the material coupled with thickness dependent optical properties, mechanical flexibility,[9] and access to valley degree of freedom highlights some of the distinct characteristics of atomically thin $MoS_2$.[10-18]

Tungsten-based dichalcogenides such as $WS_2$ and $WSe_2$ belong to the same family of layered transition metal dichalcogenides (LTMD) as $MoS_2$.[22] Mechanically exfoliated atomically thin sheets of $WS_2$ and $WSe_2$ were recently shown to exhibit high in-plane carrier mobility and electrostatic modulation of conductance similar to $MoS_2$.[23-24] Both compounds exhibit trigonal prismatic structure and are indirect gap semiconductors in the bulk form with a bandgap of ~ 1 eV as $MoS_2$. The lattice constants are nearly identical for $MoS_2$ (3.16 Å) and WS$_2$ (3.15 Å) while WSe$_2$ exhibits a slightly larger value (3.28 Å).[22] Interlayer distance is correspondingly larger for WSe$_2$ due to the large size of Se atom. General features of the WS$_2$ and WSe$_2$ band structures are similar to those of MoS$_2$ where a direct and indirect gap coexist irrespective of thickness.[25-27] Direct gap exists at the K points of the Brilloun zone between spin-orbit split valence band and doubly degenerate conduction band. On the other hand, indirect gap forms between a local conduction band minimum at a midpoint between Γ and K and valence band maximum at the Γ point. The primary difference between MoX$_2$ (where X represents chalcogen atoms) and WX$_2$ is the size of the spin-orbit splitting due to the different size of the transition metal atom.

While MoS$_2$ 2D crystals have been extensively studied recently, understanding of confinement effects on WS$_2$ and WSe$_2$ in the mono- to few-layer thickness regimes is limited. Here we report the evolution of optical absorption and PL spectra of mechanically exfoliated sheets of synthetic 2H-WS$_2$ and 2H-WSe$_2$ with thicknesses ranging betweeen 1 and 5 monolayers. The excitonic absorption and emission bands were found to gradually blueshift with decreasing number of layers due to geometrical confinement of excitons. Strong enhancement in PL efficiency is observed in monolayers indicating that they are direct gap semiconductors in agreement with the recent experimental findings[28-29] and calculations.[25-26] We show that interlayer coupling effects can be investigated by studying fine absorption and emission features and their dependence on sample thickness. We further demonstrate that the PL emission efficiency of monolayers is higher for synthetic WS$_2$ and WSe$_2$ crystals compared to natural MoS$_2$ crystals.

**Results/Discussion**

Mechanically exfoliated flakes of WS$_2$ and WSe$_2$ were deposited on quartz or SiO$_2$/Si substrates and their thickness was verified by atomic force microscopy (AFM) and optical contrast (See Supporting Information for AFM images).[30] These samples were chemically stable over 3 months in ambient conditions and showed no obvious sign of degradation according to our Raman analysis.[31] For atomically thin layers suppoted by a transparent substrate, differential reflectance provides an effective measure of absorbance. Fractional change in reflectance δR for a thin layer sample relative to the reflectance of a dielectric substrate with refractive index of $n_{subs}$ is related to the absorbance (A) of the material by[32]

$$\delta_R(\lambda) = \frac{4}{n_{subs}^2 - 1} A(\lambda).  \quad (1)$$

We assume $n_{subs}$ to be wavelength independent for the spectral range investigated in this study. Figures 1a and 1b show the differential reflectance spectra of 1-5 layer WS$_2$ and WSe$_2$, respectively. Absorption features for bulk crystals have been previously studied by Beal et al.[33] and Bromley et al.[34] General features of the peaks are in good agreement with previously reported results. All peaks exhibit gradual but distinct blueshift with decreasing sample thickness similar to early findings.[35] Excitonic absorption peaks A and B which arise from direct gap transitions at the K point are found, respectively, around 625 nm and 550 nm for WS$_2$ and around 760 nm and 600 nm for WSe$_2$ in agreement with the previous studies on bulk layers.[33-34] The energy difference between the A and B peaks, which is an indication of the strength of spin-orbit interaction, is approximately 400 meV for both WS$_2$ and WSe$_2$ in reasonable agreement with the calculations.[36] It should be noted that this value is significantly larger than the ~160 meV splitting observed for MoS$_2$.[4-5]

For WS$_2$, an additional peak, previously labeled as C, is observed around 450 nm. This peak arises from optical transitions between the density of states peaks in the valence and conduction bands.[37] The WSe$_2$ spectra show more additional features due to greater overlap of anion p orbitals with tungsten d orbitals as well as those of adjacent layers.[33-34] The absorption peaks A' and B' are believed to be excitonic in nature and arise from splitting of the ground and excited states of A and B transitions, respectively, due to inter- and intra-layer perturbation to the d electron band by the Se p orbitals.[33-34] The energy separation between A' and B' peaks are found to be ~ 400 meV consistent with this picture. It may be noted that the presence of A' and B' peaks in monolayer WSe$_2$ suggests that A-A' and B-B' splitting are mainly due to intralayer effects.

In contrast to absorption, PL spectra show remarkable dependence on sample thickness for both materials as shown in Figure 2a and b. The most notable change was a sudden increase in emission intensity when the sample was thinned to a monolayer (Figure 2c and d). Both monolayer $WS_2$ and $WSe_2$ samples exhibited distinct emission at the energy corresponding to A excitonic absorption whereas the emission intensity was dramatically reduced for multilayer samlples. While an accurage measurement of the PL quantum yield (QY) is challenging, relative changes in QY between different samples can be evaluated based on the PL and absorption spectra. Since the integrated PL intensity *I* is proportional to the product of $α(λ_{exc})$ and QY, where $α(λ_{exc})$ is the absorption coefficient at the excitation wavelength $λ_{exc}$, we can evaluate the relative magnitude of QY by comparaing $I/α(λ_{exc})$. Figure 2c and d show relative decrease in PL QY with respect to monolayer samples. For $WS_2$, PL QY drops by more than 100 folds when the thickness is increased from monolayer to bilayer and gradually diminishes with further increase in thickness. A similar jump in PL QY was observed for $WSe_2$ despite with less pronounced contrast between monolayer and bilayer samples.

For A excitonic emission from monolayer $WS_2$ and $WSe_2$, Stokes shift was typically around 20 meV and 3 meV and the corresponding full width at half maximum (FWHM) for these peaks was 75 meV and 26 meV, respectively (See Supporting Information for details). It is known that for quantum well structures Stokes shift as well as emission linewidth are an indication of interfacial quality.[38] A recent study showed that the size of Stokes shift in monolayer $MoS_2$ increases with doping concentration.[39] The narrow emission line, whose FWHM is comparable to thermal energy at room temperature, along with the small Stokes shift indicates high quality of our $WSe_2$ samples.

In addition to the A exciton peak, another peak at a longer wavelength was observed for multilayer samples of both materials. This is attributed to indirect band gap emission involving condution band minimum at a midpoint between K and Γ points and valence band maximum at the Γ point. Calculations show that for $WSe_2$ bilayers, the indirect gap emission involves valence band maximum at the K point which is nearly degenerate with the band maximum at the Γ point.[26, 40] It should be noted that the indirect gap emisision peak is virtually absent in the monolayer emission spectra, indicating that both $WS_2$ and $WSe_2$ become a direct gap semiconductor when they are thinned to a single monolayer. This observation along with the distinctly strong emission from monolayers is in good agreement with the calculated[25-27] results as well as very recent experimental findings.[28]

We observed that the decay in emission intensity with increasing thickness is more moderate for $WSe_2$ compared to $WS_2$. This is attributed to the small difference between the direct and indirect emission energies. For bulk $WSe_2$, the energy difference between the Γ and K points of the valence band was previously measured by photoemission spectroscopy to be < 80 meV.[41-42] Theoritical calculations also yield similar results[26, 40]. For bilayer $WSe_2$, the difference between the direct and indirect gaps is a measure of the energy separation between the conduction minimum at the K point and at a midpoint between Γ and K points. Based on our PL results we find this energy seperation in bilayer $WSe_2$ to be about 70 meV. In comparison, the difference between the indirect and direct band gaps in bilayer $WS_2$ is much greater (~ 300 meV). Because the population of hot electrons that transiently occupy the direct band edge at the K point is higher in bilayer $WSe_2$ due to the smaller energy difference between its direct and indirect gaps, its A exciton hot electron emission is pronounced. A similar scenario was recently reported for few-layer $MoSe_2$.[43]

For multilayer samples, the indirect gap emission intensity is comparable to or higher than that of hot electron emission from A excitons. We compared the indirect gap emission intensity of commercially available natural $MoS_2$ and our tungsten dichalcogenides grown by CVT and found that the emission intensity was distinctly higher for synthetic $WS_2$ and $WSe_2$. Radiative recombination across an indirect gap is a slow process requiring phonons with an appropriate momentum. In the presence of defects acting as nonradiative traps, nonradiative decay will dominate and suppress the PL QY. Observation of comapratively strong indirect gap emission therefore provides further evidence that our CVT-grown samples are of high quality.

Figure 3 shows absorption and PL peak energies as a funtion of the number of layers for $WS_2$ and $WSe_2$. The A and B exciton peak positions are only weakly dependent on the sample thickness while the indicret band gap energy quickly increases with decreasing thickness similar to the case of $MoS_2$. The robust direct gap at the K point is a consequence of absence of dispersion in the out-of-plane direction (or large effective mass in the out-of-plane direction). On the other hand, out-of-plane dispersion at the Γ point and along Γ–K line is greater and results in thickness dependent indirect gap. Therefore, large shift in absorption and emission peaks with flake thickness is an indication of strong out-of-plane dispersion or interlayer coupling of the electronic states. The energy of A, B, and C peaks in $WS_2$ are weakly dependent on the flake thickness indicating that the d electrons localized around the metal atoms are responsible for these optical transitions. For $WSe_2$, the shift in peaks A' and B' are more pronounced compared to that of peaks A and B suggesting that the electronic states associated with these optical transitions are more delocalized across the layers.

Photoluminescence from monolayer flakes were sufficiently strong to be imaged under fluorescence microscope. Figure 4 shows bright field optical microscope images and corresponding fluorescence images of exfoliated $WS_2$ and $WSe_2$ on quartz substrates. The fluorescence images clearly show that monolayer regions are luminescent in red consistent with the emission spectrum. Emission intensity from the adjacent bilayer regions is not sufficiently strong to be detected by the camera. The emission intensity is uniform across the monolayer region, verifying that emission is not due to anomalies arising from flake edges. Similar fluorescence images were obtained for samples deposited on $SiO_2$/Si substrates (See Supporting Information). These results demonstrate that fluorescence imaging offers a powerful tool to quickly identifying monolayers.

Ideal direct band gap semiconductors are expected to exhibit high PL QY. However, PL QY of $MoS_2$ monolayer was previously found to be extremely low (4 x $10^{-3}$).[5] The origin of the low QY is yet to be understood. Unintentional doping in the natural crystals with an estimated concentration of ~ $10^{12}$ $cm^{-2}$[1, 6, 18] may play a role in suppressing radiative recombination in addition to crystal defects. Figrue 5 shows a relative comparison of the emission intensity from the three materials. The emission efficiency of our monolayer $WS_2$ and $WSe_2$ is 20 to 40 times higher than that of monolayer $MoS_2$ obtained from a natural crystal. The emission from bilayer $WSe_2$ is also more efficient in comparison to monolayer $MoS_2$. The differences in PL QY may arise due to various intrinsic and extrinsic factors. Doping is expected to play a major role in suppressing PL in monolayers.[44-45] Since the materials studied here are unintentionally doped at different concentrations, we suspect that the observed differences mainly arise from doping. Further studies are required to elucidate the origin of the variation in PL QY reported here.

**Conclusions**

In summary, we studied the evolution of electronic structure in atomically thin $WS_2$ and $WSe_2$ sheets using differential reflectance and PL spectroscopy. Our results demonstrate that indirect to direct crossover occurs in these materials when they are thinned to a single monolayer similar to the case of $MoS_2$. Due to the excitonic recombination at direct band gap, distinct and intense PL centered at 630 nm and 750 nm is observed from monolayer $WS_2$ and $WSe_2$, respectively, which is 100 to 1000 times stronger than that in indirect-gap bulk materials. High quality of our CVT-grown samples are suggested by the distinct indirect gap emission, small Stokes shift and narrow emission linewidth. Strong interlayer coupling in $WSe_2$ is evidenced by the presence of A' and B' absorption and emission peaks and their strong thickness-dependent shift.

Our findings demonstrate the unique variations in the optical properties of 2D semiconducting crystals belonging to the family of Group 6 TMDs. Structual similarities of these materials will allow fabrication of coherent artificial crystals consisting of electronically dissimilar layers.[46-48] Monolayer $WS_2$ and $WSe_2$ with direct band gap in the visible and NIR frequency ranges, respectively, are novel building blocks for realizing unique heterostructures with tailored optoelectronic, electrocatalytic, and photocatalytic functionalities.[49]

**Method**

Synthetic crystals of 2H-$WS_2$ and 2H-$WSe_2$ were grown by chemical vapor transport (CVT) method using iodine as the transport agent. Commercial natural 2H-$MoS_2$ (SPI Supplies) was studied for comparison. The crystals were mechanically exfoliated and deposited on quartz and $SiO_2$/Si substrates for subsequent characterizations. Regions of the sample containing mono- and few-layer sheets were first identified under optical microscope and the number of layers was verified by atomic force microscope (AFM) and optical contrast. Differential reflectance measurements were performed in a backscattering geometry using a Jobin-Yvon HR800 micro-Raman system equipped with a liquid nitrogen cooled charge-coupled detector. A tungsten-halogen lamp was used as a light source. The differential reflectance is defined as 1-$R_S$/$R_Q$, where $R_S$ and $R_Q$ are the reflected light intensities from the quartz substrate with and without $WS_2$/$WSe_2$ flake samples, respectively. The light illuminating the sample was focused down to 1 μm spot using a small confocal hole. The spectra acquisition time was 15s. The intensity of the light incident on the sample was kept low (< 1 mW) in order to avoid sample damage. Photoluminescence spectra were obtained in a back scattering geometry with a 473 nm excitation laser at intensities less than 150 μW. The fluorencence images were obtained with an Olympus fluorencence microscope equipped with a Mercury lamp as the excitation light source.

**Acknowledgements**

G Eda acknowledges Singapore National Research Foundation for funding the research under NRF Research Fellowship (NRF-NRFF2011-02). PH Tan thanks the supports from NSFC under grants 10934007 and 11225421.

**Supporting Information Available**

Atomic force microscope (AFM) results, fluorescence microscope images, and additional photoluminescence spectra showing Stokes shift are presented. This material is available free of charge *via* the Internet at http://pubs.acs.org.

Figures and Figure Captions

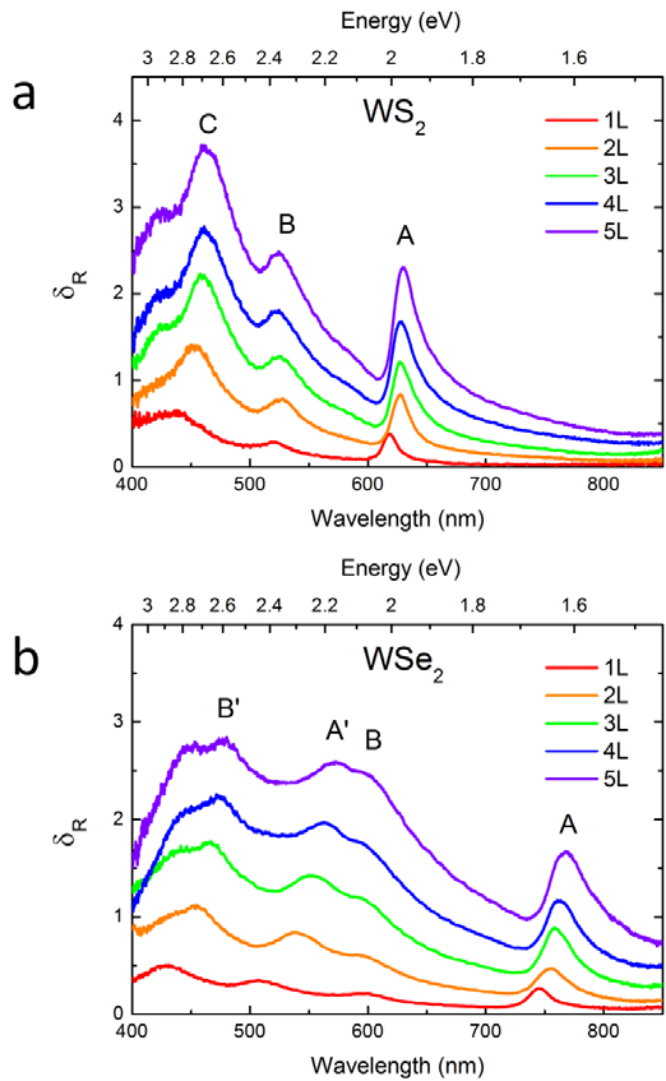

Figure 1 – Differential reflectance spectra of mechanically exfoliated (a) 2H-$WS_2$ and (b) 2H-$WSe_2$ flakes consisting of 1 to 5 layers. The peaks are labelled according to the convention proposed by Wilson and Yoffe[22].

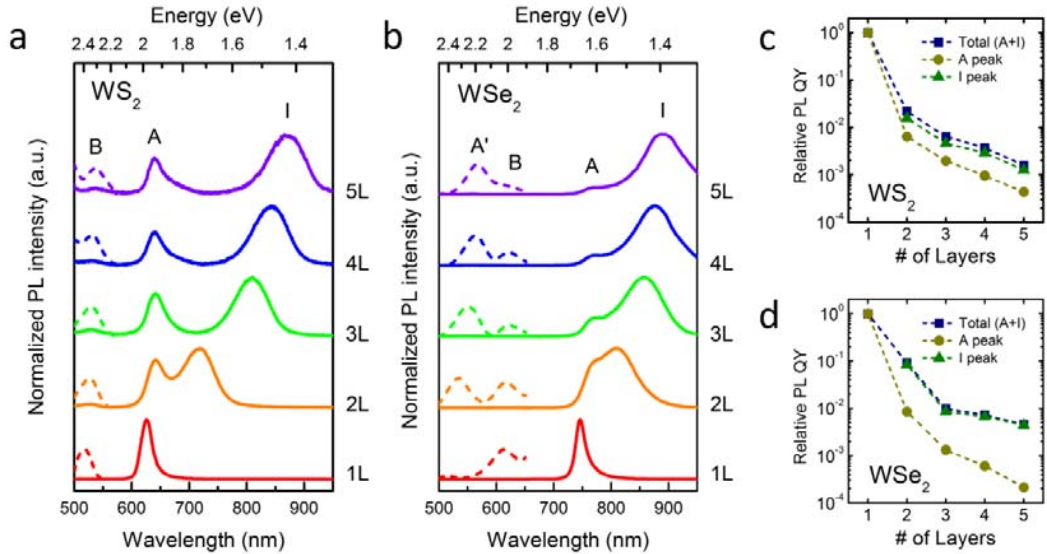

Figure 2 – Normalized photoluminescence spectra of mechanically exfoliated (a) 2H-$WS_2$ and (b) 2H-$WSe_2$ flakes consisting of 1 to 5 layers. Peak I is an indirect gap emission. Weak hot electron peaks A' and B are magnified as dashed lines for clarity. These hot electron peaks are typically 100 to 1000 times weaker than the band edge emission peaks. The total emission intensity becomes significantly weaker with increasing number of layers. Relative decay in the PL QY with the number of layers for (c) $WS_2$ and (d) $WSe_2$. The values are relative to the PL QY of a monolayer sample as discussed in the text. The plots are shown for A and I peaks and their sum (A+I).

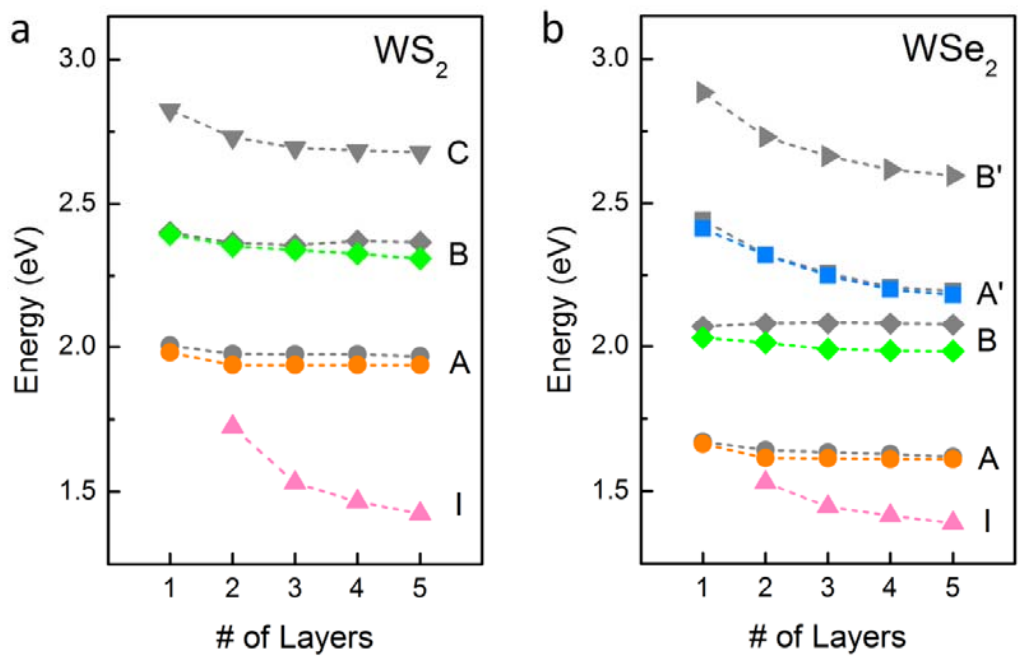

Figure 3 – Absorption (gray) and PL (color-coded) peak energies of (a) WS$_2$ and (b) WSe$_2$ flakes as a function of the number of layers. The letters I, A, B, A', B', and C refer to the peaks labeled in Figure 1 and 2.

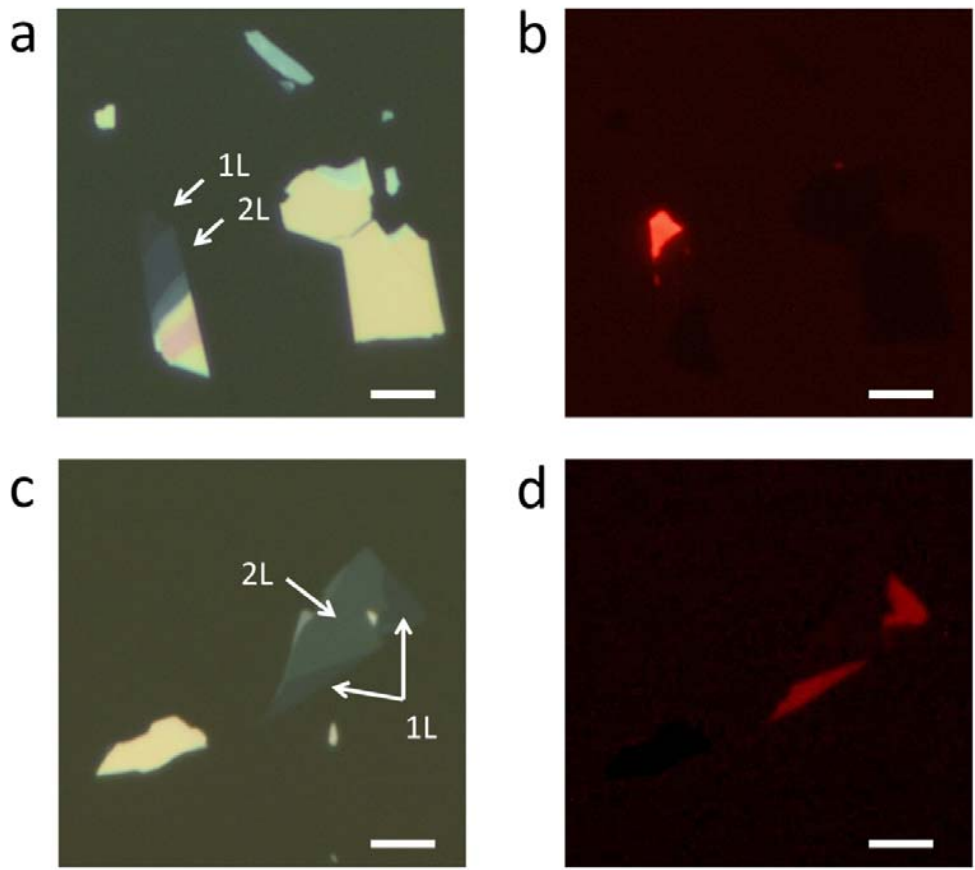

Figure 4 – Bright field optical images of mechanically exfoliated (a) WS$_2$ and (c) WSe$_2$ and their corresponding fluorescence images (b, d). The scale bar is 5 µm. The fluoresence images were obtained with excitation wavelength of 510-550 nm. Due to low efficiency of the CCD cameara in the NIR range, fluorescence image contrast is artificially enhanced for WSe$_2$. The original image can be found in the Supporting Information.

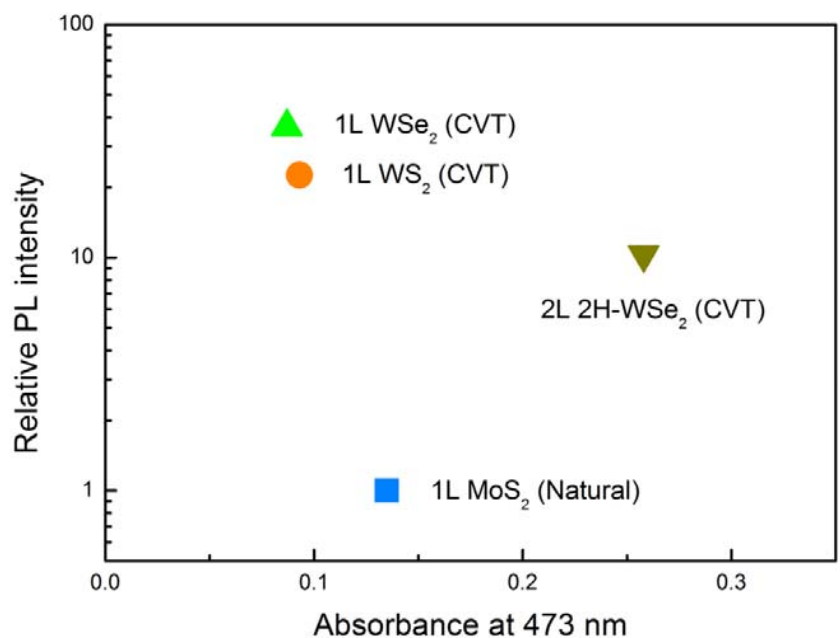

Figure 5 – Relative integrated PL intensity of $WS_2$ and $WSe_2$ monolayers and $WSe_2$ bilayer with respect to that of a $MoS_2$ monolayer plotted against absorbance at the excitation wavelength of 473 nm. The PL emission spectra were integrated over 520 to 950 nm. Absorbance is obtained from differential reflectance of samples on quartz based on Eqn 1. The absorbance of $MoS_2$ is obtained from Ref [5]. Note that the comparison is made between exfoliated samples from synthetic (*i.e.* CVT-grown) $WS_2$/$WSe_2$ and natural $MoS_2$ (SPI supplies) crystals.

# Supporting Information

# Evolution of electronic structure in atomically thin sheets of WS$_2$ and WSe$_2$

*Weijie Zhao[a,c,#], Zohreh Ghorannevis[a,c,#], Leiqiang Chu[a,c], Minglin Toh[d], Christian Kloc[d], Ping Heng Tan[e], Goki Eda[a,b,c,*]*

[a] Department of Physics, National University of Singapore, 2 Science Drive 3, Singapore 117542

[b] Department of Chemistry, National University of Singapore, 3 Science Drive 3, Singapore 117543

[c] Graphene Research Centre, National University of Singapore, 6 Science Drive 2, Singapore 117546

[d] School of Materials Science and Engineering, Nanyang Technological University, N4.1 Nanyang Avenue, Singapore 639798

[e] State Key Laboratory for Superlattices and Microstructures, Institute of Semiconductors, Chinese Academy of Sciences, Beijing, China, 100083

[*] E-mail: g.eda@nus.edu.sg

[#] These authors contributed equally to this work.

1. AFM images

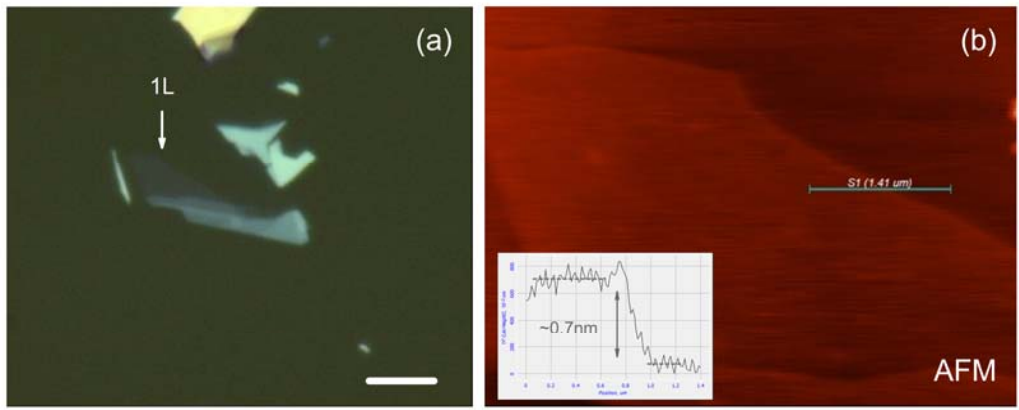

Figure S1 | Optical (a) and AFM (b) images of monolayer $WS_2$ on a quartz substrate. The scale bar in (a) is 5 μm.

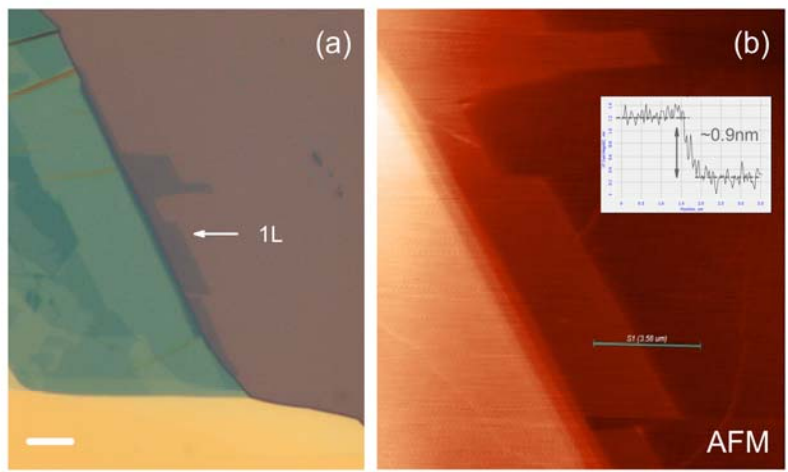

Figure S2 | Optical (a) and AFM (b) images of monolayer $WSe_2$ on a $SiO_2/Si$ substrate. The scale bar in (a) is 5 μm.

2. Fluorescence images

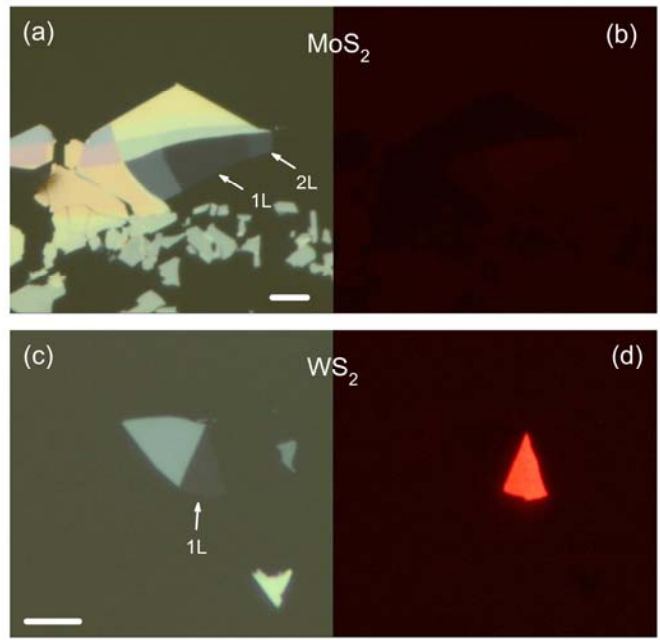

Figure S3 | Optical (a) and fluorescence (b) images of monolayer $MoS_2$. Optical (c) and fluorescence (d) images of monolayer $WS_2$. Both samples were deposited on a quartz substrate and the fluorescence images were taken under identical conditions. The scale bar in (a) and (c) is 5 μm.

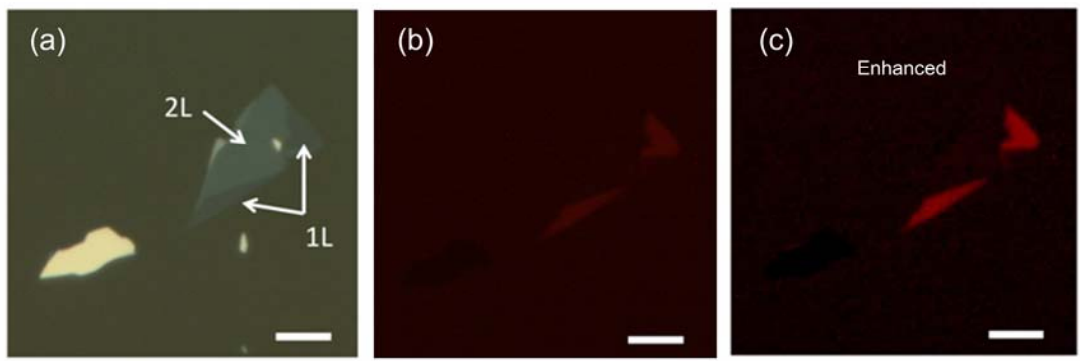

Figure S4 | Optical (a) and fluorescence (b) images of monolayer $WSe_2$ on quartz substrate. The fluorescence in (b) looks weak because of the low responsivity of the CCD of the fluorescence microscope in the near infrared range. Constrast

enhanced fluorescence image (c) is shown in Figure 4d of the main text. The scale bar is 5 µm.

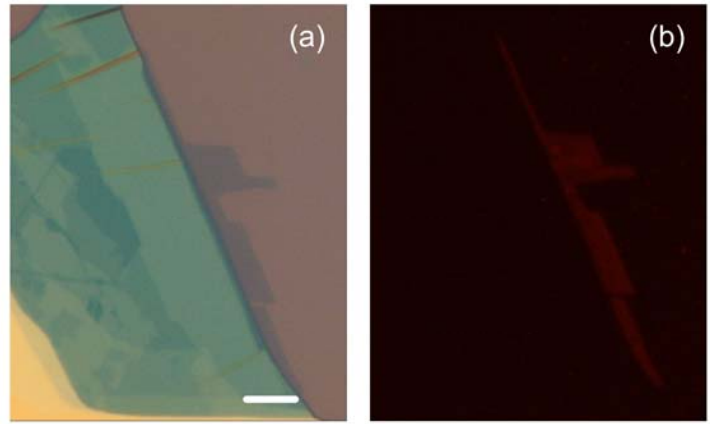

Figure S5 | Optical (a) and fluorescence (b) images of monolayer $WSe_2$ on a $SiO_2$/Si substrate. The scale bar in (a) is 5 µm.

3. Stokes shift

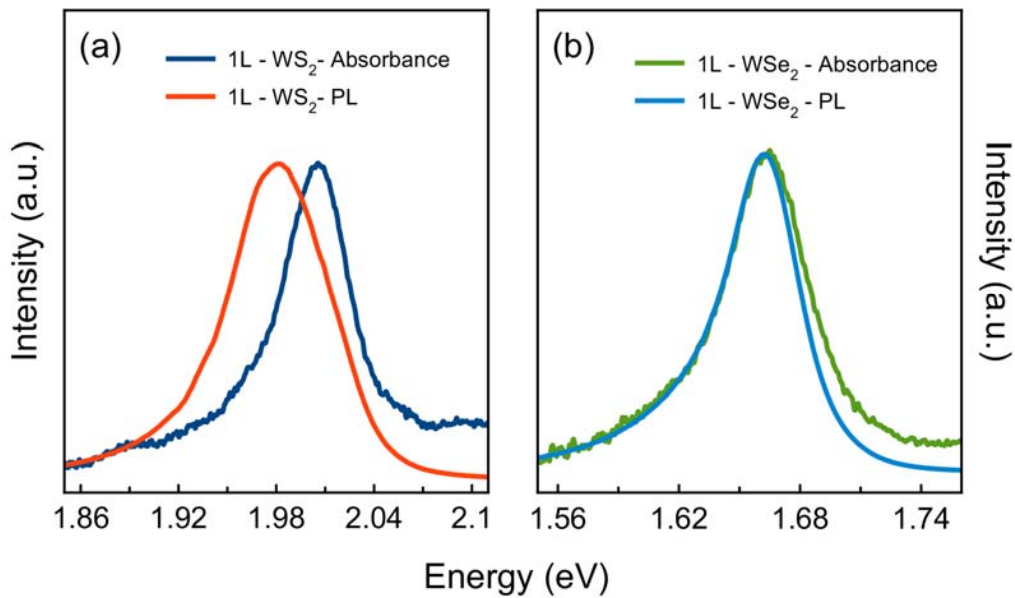

Figure S6 | Stokes shift of monolayer $WS_2$ (a) and $WSe_2$ (b). The absorbance spectra are equivalent to the differential reflectance spectra presented in the main text. The spectra are normalized and overlaid for clarity.